\documentclass[12pt,a4paper]{article}
\usepackage{color}
\usepackage[all]{xy}
\usepackage{booktabs}
\usepackage{authblk}
\usepackage[T1]{fontenc}
\usepackage{lmodern}
\usepackage[utf8]{inputenc}
\usepackage[english]{babel}
\usepackage{caption}
\usepackage{subfig}

\usepackage{epsfig}
\usepackage{graphicx}
\usepackage{amsmath}
\usepackage[titletoc,title]{appendix}
\usepackage{amsfonts}
\usepackage{amssymb}
\usepackage{indentfirst}

\title{{\bf{Dynamics of a perfect fluid through velocity potentials with aplication in quantum cosmology}}}

\author[1]{F. G. Alvarenga\thanks{e-mail: \texttt{flavio.alvarenga@ufes.br}}}
\author[2]{R. Fracalossi\thanks{e-mail: \texttt{rfracalossi@gmail.com}}}
\author[1]{R. G. Furtado\thanks{e-mail: \texttt{raphael.furtado@ufes.br}}}
\author[2]{S. V. B. Gon\c{c}alves\thanks{e-mail: \texttt{sergio.vitorino@pq.cnpq.br}}}
\affil[1]{Departamento de Ci\^encias Naturais, CEUNES, Universidade Federal do Esp\' \i rito Santo, CEP 29933-415, S\~ao Mateus, ES,
Brazil.}
\affil[2]{Departamento de F\' \i sica, Centro de Ci\^encias Exatas, Universidade Federal do Esp\' \i rito Santo, CEP 29075-910, Vit\'oria, ES, Brazil.}

\date{}
\begin{document}
\maketitle

\begin{abstract}
We review the Eulerian description of hydrodynamics using Seliger-Whitham's formalism (in classical case) and Schutz's formalism (in relativistic case). In these formalisms, the velocity field of a perfect fluid is described by scalar potentials. With this we can obtain the evolution equations of the fluid and its Hamiltonian. In the scenario of quantum cosmology the Schutz's formalism makes it possible to introduce phenomenologically a time variable in minisuperspace models.
\end{abstract}

PACS number(s): 98.80.-k, 98.80.Cq, 98.80.Qc

\section{Introduction}

The study of hydrodynamics always been present in physics, both in its classical context as in modern physics \cite{landau}. Its theoretical origins date back to the eighteenth century with Daniel Bernoulli's works comprising two older areas, namely the hydrostatic and hydraulic. With the hydrostatic many contributions came from the Greeks and Romans. Archimedes and your principle and Pascal also contributed to the study of incompressible flow (the divergence of the flow velocity is zero) and at rest. Hydraulic already covered a more practical area involving fundamental engineering and applied sciences. In general, the interest in hydraulic was related to the technology involving irrigation methods, events in agriculture and grinding grains. Despite several phenomenological studies, the  hydrodynamics reached a high level in the eighteenth century. In addition to Bernoulli, contributions were made by many scientists as d'Alembert, Euler, Navier, Stokes, Helmholtz and Kirchhoff among others.
\par
The study of fluid flow was developed before the conclusion of a great discussion about if the matter was made up of discrete particles or could be considered as a continuous, infinitely divisible, where the matter was seen as a whole. In the case of hydrodynamics, we must address the phenomena of the macroscopic point of view, with its defined properties at each point in space, and not considering microscopic elements. Thus, in a large number of situations it is possible to treat a fluid as a solid substance being identified with a set of points in $R_N$, where $N$ is related to the dimension to which we are working.
\par
To define a fluid we can use two specific descriptions: the Eulerian description and the Lagrangian description. Consider a discrete parcel or volume of the fluid carrying their own physical properties that in turn vary with time. The description of the fluid as a whole, made following the evolution of each of the parcels moving in time is called the Lagrangian description. The basic properties of a given parcel of the fluid can be represented in the Lagrangian description as $\rho(t)$, $p(t)$, $\vec v(t)$, etc. The Lagrangian method is often used in the description of experiments and in numerical simulations.
\par
If, on the other hand, we consider the evolution of the physical properties at each point in space, instead of following the evolution of the parcels of fluid, we have the Eulerian description. It can be regarded as a field description in the sense that every point in space have certain physical properties that can be studied and analyzed. Thus, the properties of the fluid flow at a specific position will depend on the spatial position and temporal coordinate. From a mathematical point of view, we can represent the properties of the fluid as $\rho(\vec x, t)$, $\vec v(\vec x, t)$, $p(\vec x, t)$, etc. In general, to be mathematically simpler to apply we mainly use the Eulerian description.
\par
The formalism that we review in this article it has an ancient origin. Already in 1859, Clebsch \cite{Clebsch} showed that the velocity field of a fluid (in fact, any vector field) could be described with the use of three scalar potentials in the form
\begin{displaymath}
\vec{v}=\vec{\nabla}\phi+\alpha\vec{\nabla}\beta\quad.
\end{displaymath}
However, this decomposition had a disadvantage: the potentials  $\phi$, $\alpha$ and
$\beta$  had no clear physical interpretations because there were no individual evolution equations for them.
\par
Seliger and Whitham \cite{Seliger}, in order to avoid this difficulty, made a decomposition of the velocity field in terms of five scalar potentials, rather than the minimum number of Clebsch. Using this technique they have obtained the evolution equations of a classical fluid in Eulerian description through a variational principle. In this formalism, the Lagrangian density is extremely simple, being only the fluid pressure.
\par
The idea of using a hydrodynamic analogy in other areas of physics is quite common and the list is big. We find, for example, these analogies in studies involving the following areas: plasma physics, optics, many problems involving electromagnetism, turbulence, solid state physics in liquid crystal studies, elastic media studies, acoustics, atmospheric physics, geophysics, magnetism, astrophysics, cosmology, etc.
\par
In the case of cosmology, the analogy is of fundamental importance because of the complexity of the subject. The matter that fills the universe as a whole is regarded as a fluid that expands with galaxy clusters representing the particles of this fluid. The simplest case is one where we considered the fluid behaving like a perfect fluid with a state equation of type $p = \alpha\rho$. In more elaborate models we can consider torsion, shear, etc. In the case of quantum cosmology the Schutz's variational formalism, \cite{Schutz1, Schutz2} is a relativistic generalization of the work of Seliger-Whitham and Whitham, now making use of six velocity potentials (one more than in the Newtonian case because now we use four-vectors like four-velocity) and including the effects of the gravitational field, from the point of view of general relativity.
\par
Here in this work we propose to do a review of the work of Seliger-Whitham and Schutz, showing their similarities and differences. We also make a discussion of Schutz's formalism applied to quantum cosmology.
\par
The outline of the paper is as follows. In Section \ref{SWF}, we introduce the Seliger-Whitham formalism. We begin, in Subsection \ref{CtA} constructing the action and we derive the evolution equation of the hydrodynamical system. Then, in Subsection \ref{TOE}, we obtain the evolution equations through the Hamilton principle. In Section \ref{TSF}, we show how to construct the Schutz’s formalism with the relativistic velocity potentials representation. For starters, in Subsection \ref{TRF}, we introduce the basic relativistic formalism. In Section \ref{QCSF}, we show how we can use the Schutz's formalism in quantum cosmology. Finally, in Section \ref{Con}, we show our final words about this subject. In Appendix \ref{app} we provides a elementary study of the thermodynamics of one perfect fluid.

\section{Seliger-Whitham Formalism}
\label{SWF}

In 1968, Seliger and Whitham \cite{Seliger} released an important article describing the hydrodynamics, plasma dynamics and elasticity through the variational principle. For this they use the representation by velocity potentials to write the electromagnetism field equations and fluid dynamics in Eulerian description. The latter case is what interest us here in this article. This construction is presented below.

\subsection{Constructing the Action}
\label{CtA}

In the Eulerian description of fluid mechanics{\footnote{as Reference \cite{Schutz1}, "{\it{History has mercilessly given us half a dozen different uses for the names of Lagrange and Euler. The adjectives Lagrangian and Eulerian refer, respectively, to observers comoving with the fluid or fixed with respect to some arbitrary reference frame through which the fluid flows. The functional whose integral is extremized in a variational principle is the Lagrangian density. Finally the equations that express the extremal conditions are the Euler-Lagrange equations. Because we wish to emphasize the Eulerian nature of the velocity potentials, we shall henceforth speak of their evolution equations rather than of their equations of motion.}"}}, the independent variables are the position (fixed) $\vec{X}$ and the time $t$. The Eulerian velocity $\vec{v}(\vec{X}, t)$ is the velocity of the parcel \footnote{ Is called parcel the fluid element with non-null infinitesimal volume and surface area.} of fluid passing by $\vec{X}$ at time $t$. All other dependent quantities are analogously defined. The evolution equation is described as
\begin{equation}\label{euler1}
\rho\left(\frac{\partial V^i}{\partial t} + V^j \frac{\partial V^i}{\partial X^j}\right)=-\frac{\partial p}{\partial X^i}\quad.
\end{equation}
\par
In addition, we will impose as constraints the conservation of mass
\begin{equation}\label{massa}
\frac{\partial\rho}{\partial t}+\frac{\partial}{\partial X^j}(\rho V^j) = 0\quad,
\end{equation}
and the conservation of specific entropy (the entropy per unit mass) 
\begin{equation}\label{entropia}
\frac{\partial s}{\partial t}+V^j\frac{\partial s}{\partial X^j} = 0\quad.
\end{equation}
The constraint ({\ref{entropia}}) implies the absence of heat flow through the walls of the parcel.
\par
To these results we add the equation of state
\begin{equation}\label{estado}
p = p(\rho,s)\quad.
\end{equation}
\par
The central issue here is: how to get the equations (\ref{euler1}), (\ref{massa}) and (\ref{entropia}) from a variational principle? The idea would use the action \footnote{It is appropriate to modify the traditional Lagrangian density, replacing the potential energy of the system by the internal energy as \cite{herivel}.}
\begin{eqnarray}
\mathcal{S}_{\varepsilon} &=& \int dt \int d^3X \bigg\{ \left[\frac{1}{2} \rho V^iV_i -\rho\varepsilon\right] \nonumber\\
&+&\phi \left[\frac{\partial\rho}{\partial t}+\frac{\partial}{\partial X^j}(\rho V^j)\right]+\theta\left[\frac{\partial s}{\partial t}+V^j\frac{\partial s}{\partial X^j}\right]\bigg\}\quad,\label{lageul1}
\end{eqnarray}
where $\rho(X^i, t)$ is the density of the fluid, $V^i (X^i, t)$ the fluid velocity, $\varepsilon(X^i, t)$ is the internal energy per unit of mass of the fluid passing through the
point $X^i$ at the instant $t$ and $\phi$ and $\theta$ are Lagrange multipliers.
\par
By varying the above action with respect to $V^i$, we obtain the following equation
\begin{equation}\label{problema}
V_i=\frac{\partial \phi}{\partial X^i}+ s\frac{\partial \theta}{\partial X^i}\quad.
\end{equation}
\par
However, as discussed in Reference \cite{Seliger}, this result is not as general as we would desire. If the entropy density $s$ is constant, the velocity field $V^i$, equation (\ref{problema}), can be written as
\begin{displaymath}
V_i=\frac{\partial (\phi + s\theta)}{\partial X^i}\quad,
\end{displaymath}
and the conclusion obtained is that the above equation expresses a velocity field as the gradient of a scalar, i.e., $V^i = \partial_i\Phi^i$, where $\Phi = \phi + s\theta$. But the curl of a vector field written as a gradient of a scalar function is zero. In our case we have $\varepsilon_{ijk}\partial_j{V^k} = 0$. This is called potential flow or irrotational flow, a subset of  flow of an ideal fluid. As we want that rotational fluids could be described even when the entropy per unit mass does not vary spatially, it is necessary to change (\ref{lageul1}).  To resolve this issue, the Reference \cite{Lin} proposed the inclusion of a new constraint (and, therefore, a new free parameter) in action expression: the conservation of the starting position during the flow
\begin{equation}
\frac{\partial \alpha^i}{\partial t}+ V^j\frac{\partial\alpha^i}{\partial X^j}=0\quad.
\end{equation}
\par
Including this remaining of the Lagrangian description, we now have three new constraint's equations. However, also as discussed in Reference \cite{Seliger}, for our purposes, we consider just a single component, which will identify for $\alpha$ only.
\par
Thus, the equation (\ref{lageul1}) becomes
\begin{eqnarray}
\mathcal{S}_\varepsilon = \int dt \int d^3X \bigg\{ \left[\frac{1}{2} \rho V^i V_i -\rho\varepsilon\right] +\phi \left[\frac{\partial\rho}{\partial t}+\frac{\partial}{\partial X^j}(\rho V^j)\right]\nonumber\\
+\theta\left[\frac{\partial s}{\partial t}+V^j\frac{\partial s}{\partial X^j}\right]\bigg\}+ \beta\left[\frac{\partial \alpha^i}{\partial t}+ V^j\frac{\partial\alpha^i}{\partial X^j}\right]\quad.\label{lageul2}
\end{eqnarray}
By varying the action, equation (\ref{lageul2}), with respect to $V^i$, we get
\begin{equation}\label{potvel}
V_i=\frac{\partial \phi}{\partial X^i}+ s\frac{\partial \theta}{\partial X^i}+\alpha\frac{\partial \beta}{\partial X^i}\quad,
\end{equation}
so it is possible to describe the vorticity independently of entropy gradients. Still, by varying with respect to  $\rho$, $s$ and $\alpha$ we obtain,  respectively
\begin{equation}\label{rho1}
\frac{1}{2} V^iV_i-\frac{\partial}{\partial\rho}(\rho\varepsilon)=\left(\frac{\partial\phi}{\partial t}+V^j\frac{\partial\phi}{\partial X^j}\right)+
s\left(\frac{\partial\theta}{\partial t}+V^j\frac{\partial\theta}{\partial X^j}\right)+\alpha\left(\frac{\partial\beta}{\partial t}+V^j\frac{\partial\beta}{\partial X^j}\right)\quad,
\end{equation}
\begin{equation}
\frac{\partial \varepsilon}{\partial s}=-\left(\frac{\partial\phi}{\partial t}+V^j\frac{\partial \theta}{\partial X^j}\right)\quad,
\end{equation}
and
\begin{equation}
\frac{\partial \phi}{\partial t}+ V^j\frac{\partial \theta}{\partial X^j}=0\quad.
\end{equation}
\par
Integrating by parts the equation (\ref{lageul2}) and discarding surface terms, we get
\begin{equation}\label{por partes}
{\cal{L}}=\frac{1}{2}\rho v^2 -\rho \varepsilon - \rho\frac{D \phi}{D t}- \rho s\frac{D \theta}{D t}-\rho \alpha\frac{D \beta}{D t}\quad,
\end{equation}
where we have used the usual notation of material derivative, also called substantial derivative, that describes the time rate of change of some physical quantity like energy, temperature or momentum at a fixed location, seen by an observer moving along with the flow with velocity $V^j$ and given by
\begin{equation}\label{material}
\frac{D}{Dt}\equiv \frac{\partial}{\partial t} + V^j\frac{\partial}{\partial x^j}\quad.
\end{equation}
In general, the material derivative can serve as a link between Eulerian and Lagrangian descriptions of continuum deformation
\par
To simplify the expression of the Lagrangian density, we use the first law of thermodynamics
\begin{equation}\label {first}
d\varepsilon=Tds - pd\left(\frac{1}{\rho}\right)\quad,
\end{equation}
where $d\varepsilon$ is the internal amount of energy per unit rest mass and $\rho$ is the density of rest mass.
\par
Consequently, we have
\begin{eqnarray}
\label{ental}
\frac{\partial(\rho \varepsilon)}{\partial\rho} &=&\varepsilon + \frac{p}{\rho}\equiv \mu\quad,\\
\frac{\partial\varepsilon}{\partial s} &=& T\quad,
\end{eqnarray}
where $\mu(\rho,s)$ is the enthalpy per unit mass, which is the sum of the internal energy of the system plus the product of the pressure of the gas in the system times the volume of the system, and $T(\rho,s)$ is temperature.
\par
Substituting the equation (\ref{ental}) into equation (\ref{rho1}) we have the next result 
\begin{equation}\label{entalpia}
\mu = \frac{1}{2}  v^2 - \frac{D \phi}{D t}- s\frac{D \theta}{D t}-\alpha\frac{D \beta}{D t}\quad.
\end{equation}
\par
Finally, substituting equation (\ref{entalpia}) into equation (\ref{por partes}), and using the equation (\ref{ental}) we get this extremely simple form to the Lagrangian density, equation (\ref{por partes}), and the new corresponding action
\begin{equation}\label{simples}
{\cal{L}}=\rho \mu - \rho \varepsilon= p\quad \rightarrow\quad \mathcal{S} = \int d^3x\, p\quad.
\end{equation}

\subsection{The old equations in the new formalism}
\label{TOE}

We will now use the previous Lagrangian density, equation (\ref{simples}), to obtain the evolution equations through the Hamilton principle. For this, we need to vary the following expression
\begin{equation}
\mathcal{S} = \iint p(\mu,s) d^3\!x\, dt\quad.
\end{equation}
\par
By equations (\ref{first}) and (\ref{ental}), we see that
\begin{equation}
dp=\rho d\mu-\rho T ds\quad.
\label{press}
\end{equation}
\par
Introducing {\it a priori} the velocity potentials representation, equation (\ref{potvel})
\begin{equation}
V_i=\partial_i \phi+s\partial_i\theta+\alpha\partial_i\beta\quad,
\end{equation}
together with the equation (\ref{entalpia}), conveniently rewritten in the form
\begin{equation}
\mu=-\frac{\partial\phi}{\partial t}-s\frac{\partial\theta}{\partial t}-\alpha\frac{\partial \beta}{\partial t}-\frac{1}{2}\left(\nabla\phi+s\nabla\theta+\alpha\nabla\beta\right)^2\quad,
\end{equation}
we obtain
\begin{eqnarray}
\frac{D\rho}{D t}&=&-\rho\nabla\cdot\boldsymbol{v}\label{rho}\quad,\\
\nonumber\\
\frac{D\theta}{D t}&=&-T\label{eta}\quad,\\\nonumber\\
\frac{D s}{D t}&=&0\label{S}\quad,\\\nonumber\\
\frac{D\beta}{D t}&=&0\label{beta}\quad,\\\nonumber\\
\frac{D\alpha}{D t}&=&0\label{alpha}\quad.
\end{eqnarray}
The equations in the new formalism (\ref{rho})-(\ref{alpha}) have been proved to be equivalent to the traditional ones (\ref{euler1})-(\ref{entropia}), in the sense that all the solutions of the traditional equations are included in the solutions of the new ones.

In order to obtain the hamiltonian density, we will follow \cite{Saarloos}. The authors calls our attention to the fact that, contrary to the usual case, we obtain the same number of equations that using the lagrangian formalism. This occours because the lagrangian equations are already of first order in time.
We have only three non-zero conjugated momenta. They are
\begin{equation}
\label{ham1}
\Pi_{\phi}\equiv\frac{\partial \cal{L}}{\partial \dot{\phi}}= - \rho,
\end{equation}
\begin{equation}
\label{ham2}
\Pi_{\beta}\equiv\frac{\partial \cal{L}}{\partial \dot{\beta}}= - \rho\alpha,
\end{equation}
\begin{equation}
\label{ham3}
\Pi_{\theta}\equiv\frac{\partial \cal{L}}{\partial \dot{\theta}}= - \rho s.
\end{equation}

The hamiltonian density is simply
\begin{equation}
{\cal{H}}=\dot{\phi}\Pi_{\phi}+\dot{\beta}\Pi_{\beta}+\dot{\theta}\Pi_{\theta}-{\cal{L}},
\end{equation}
and, using (\ref{ham1}) - (\ref{ham3}), we obtain
\begin{equation}
{\cal{H}}=-\frac{1}{2 \Pi_{\phi}}\left(\Pi_{\phi}\nabla\phi+\Pi_{\beta}\nabla\beta+\Pi_{\theta}\nabla\theta\right)^2-\Pi_{\phi}\varepsilon(\Pi_{\phi},\Pi_{\theta}).
\end{equation}

\section{The Schutz's formalism: the relativistic velocity potentials representation}
\label{TSF}

Between 1970 and 1971, B. F. Schutz published two seminal articles \cite{Schutz1}-\cite{Schutz2} with the aim of analyzing the behavior of a perfect fluid in a relativistic version. As in the Seliger-Whitham formalism, this version also writes the hydrodynamic equations of perfect fluid through the velocity potentials. In this way, we can understand this relativistic version as a generalization of treatment provided through the Seliger-Whitham formalism presented in the previous section, although there is not a formal demonstration of reduction of relativistic version to the  Newtonian version, as happens in general relativity and quantum mechanics in relation to Newtonian mechanics.
\par
The Schutz's articles describe the perfect fluid in Eulerian form in terms of six velocity potentials, where five are scalar fields and the sixth is the entropy $s$. When we vary the action in relation to the scalar fields we obtain the evolution equations of the velocity potentials and when we vary the action in terms of the metric tensor, we obtain the Einstein equation that represent this fluid.
\par
The Hamiltonian description of fluid is also obtained by Schutz. This analysis is fundamental to the quantization process of physical systems that will be seen more forward (considering here only the first quantization or canonical quantization or quantization of Dirac), associating classical observables with quantum operators. Through the Arnowitt-Deser-Misner (hereafter referred to as ADM) canonical theory we split the four-dimensional spacetime in an appropriate (3 + 1)-dimensional spacetime considering an observer at rest in a hypersurface with constant coordinate time. The description of the fluid is made in terms of two independent sectors, one related to energy density and another related to the fluid.

\subsection{The relativistic formalism}
\label{TRF}

The relativistic one component perfect fluid is represented by a equation of state of type $p = p(\mu, s)$ and the energy-moment tensor written as
\begin{align}
\nonumber T^{\mu\nu}&=(\rho+p)U^{\mu}U^{\nu}+pg^{\mu\nu}\\
&=\rho_0 \mu\,U^{\mu}U^{\nu}+pg^{\mu\nu}\quad,\label{129}
\end{align}
where $\mu = (\rho + p)/\rho_0$ is the specific inertial mass, also known as enthalpy, as defined by Schutz in References \cite{Schutz1, Schutz2}. At this point it is interesting to make clear that there is no standardized notation in the literature regarding the relativistic hydrodynamics and also with the Newtonian interpretation developed by Seliger-Whitham \cite {Seliger}. We present in Table \ref{table:notation} a comparison of the notations used in this article, similar to other tables shown in relativistic hydrodynamics books \cite{HR}. This discussion will be examined again in the Appendix \ref{app}.
\begin{table}[ht]
\caption{Hydrodinamic notations.}
\centering
\begin{tabular}{c c c }
\hline \hline
Quantity/Reference & Seliger \cite{Seliger} & Schutz \cite{Schutz1}\\  [0.5ex]
\hline
rest-mass density & $\rho$ & $\rho_0$ \\
mass-energy density & $\varepsilon$ & $\rho = \rho_0 (1 + \Pi)$ \\
Enthalpy & $\mu = \varepsilon + \frac{p}{\rho}$ & $\mu = 1 + \Pi + \frac{p}{\rho_0} = \frac{\rho + p}{\rho_0}$ \\
\hline
\end{tabular}
\label{table:notation}
\end{table}
\par
Because it is a perfect fluid, the energy-moment tensor shows no viscosity or heat conduction and his signature on a locally comoving inertial reference is $(\rho,p,p,p)$.
\par
The condition for the conservation of the baryons number is expressed as
\begin{equation}
(\rho_0\,U^{\mu})_{;\mu}=0\quad.\label{130}
\end{equation}
Classically, this result corresponds to mass conservation, expressed by the equation (\ref{massa}). Furthermore, the normalization condition of four-velocity is
\begin{equation}
U^{\nu}U_{\nu}=-1\;,\label{131}
\end{equation}
which leads to
\begin{equation}
U^{\nu}U_{\nu\;;\sigma}=0\quad.\label{132}
\end{equation}
\par
Considering the moment-energy tensor with zero divergence, i.e.,
\begin{equation}
T^{\mu\nu}\,_{;\nu}=0\quad,\label{133}
\end{equation}
the evolution equations can be expressed in the form of conservation laws.
\par We will follow the references \cite{Schutz1} and \cite{Tovar} to obtain the equations of hydrodynamics of relativistic fluid.
\par
On the constraints imposed by the eq. (\ref{130}) and eq. (\ref{131}), the four equations in (\ref{133}) will determine the movement of a fluid for which is known the equation of state. To get a clear interpretation of eq. (\ref{133}), the perpendicular and parallel components to the four-velocity should be separated. For the parallel component, we write
\begin{equation}
U_{\mu}T^{\mu\nu}\,_{;\nu} = U_{\mu}[\rho_0 \mu\,U^{\mu}U^{\nu}+pg^{\mu\nu}]\,_{;\nu} = U^{\nu}p_{,\nu}-\rho_0\,U^{\nu} \mu_{,\nu} = 0\quad,
\label{entrop}
\end{equation}
where we used above the equations (\ref{130}), (\ref{131}) and (\ref{132}). By $p_{,\nu} = \rho_0 \mu_{,\nu} - \rho_0~T s_{,\nu}$, the relativistic version of the equation (\ref{press}), the expression (\ref{entrop}) becomes
\begin{equation}
\rho_0TU^{\nu}s_{,\nu}=0\quad,\label{136}
\end{equation}
i.e., the motion of a relativistic perfect fluid preserves the entropy per baryons. Classically this result matches with specific entropy conservation, equation (\ref{entropia}).
\par
To make explicit the perpendicular component, we define the projection tensor
\begin{equation}
P^{\sigma}\,_{\mu}=\delta^{\sigma}\,_{\mu}+U^{\sigma}U_{\mu}\quad.\label{137}
\end{equation}
Under the action of the above projection tensor the equation (\ref{133}) become
\begin{displaymath}
P^{\sigma}\,_{\mu}T^{\mu\nu}\,_{;\nu}=0\quad,
\end{displaymath}
and by using the equations (\ref{130}), (\ref{131}) and (\ref{132}), it is possible to write
\begin{equation}
-P_{\sigma}\,^{\nu}p_{,\nu}=\mu \rho_0\,U_{\sigma;\nu}U^{\nu}\quad.
\label{pressao}
\end{equation}
If we take $U^{\nu}=\delta^{\nu}\,_0$, we find the following expression
\begin{equation}
-\vec{\nabla}p=(\rho+p)\frac{d\vec{v}}{d\tau}\quad,\label{140}
\end{equation}
where $\vec{v}$ is the three velocity and $\tau$ is the proper time. Thus, we verify that the expression (\ref{pressao}) is the relativistic version of (\ref{euler1})

Using again the Pfaff theorem and considering the four-velocity as a four-dimensional function of the spacetime coordinates, we see that there are four scalar functions ($A, B, C, D$) that describes the four-velocity, in the form
\begin{displaymath}
U_{\nu}=AB_{,\nu}+CD_{,\nu}\quad.
\end{displaymath}
\par
Schutz extended the Seliger-Whitham idea for the relativistic formalism with introduction of two new potential, each one with its evolution equation. Thus, the four-velocity takes the form
\begin{equation}
U_\nu=\mu^{-1}(\phi_{,\nu}+\alpha\beta_{,\nu}+\theta\,s_{,\nu})\quad.\label{141}
\end{equation}
Thus, $s$ is the entropy per baryon and $\mu$ the enthalpy. This is a generalization of equation (\ref{potvel}) with the introduction of the enthalpy as a new potential and the replacement of the term $-s\theta,_{\nu}$ for $\theta s,_{\nu}$. 
\par
The potentials $\alpha$ and $\beta$ are associated with rotational movement and must be null in theories that describe homogeneous systems.
\par
From the imposing of equation (\ref{131}), we have
\begin{equation}
\mu^2=-g^{\sigma\nu}(\phi_{,\sigma}+\alpha\beta_{,\sigma}+\theta\,s_{,\sigma})
(\phi_{,\nu}+\alpha\beta_{,\nu}+\theta\,s_{,\nu})\quad,\label{142}
\end{equation}
that is, the enthalpy is a function of the other velocity potentials, similarly as equation (\ref{entalpia}).
\par
For the formulation be useful, in conformity with the four-velocity shown by equation (\ref{141}), it is necessary that the known equations of a perfect fluid can be reproduced using an appropriate action. As mentioned previously, the Schutz's representation is associated with a variational Lagrangian principle that generalizes the equation (\ref{simples}) to the general relativistic case as follows
\begin{equation}
\mathcal{L}=\sqrt{-g}(R+16 \pi p)\quad,\label{143}
\end{equation}
where $R$ is the curvature scalar of the time-space and $p$ is the pressure of the fluid.
\par
Consider following action
\begin{equation}
\mathcal{S}=\int\,d^4x\sqrt{-g}(R+16\pi\,p)\quad.\label{144}
\end{equation}
The calculation of the variation of the action above is obtained from
\begin{displaymath}
\delta(\sqrt{-g}R)=\left(R_{\sigma\nu}-\frac{1}{2}g_{\sigma\nu}R\right)\sqrt{-g} \,\delta{g^{\sigma\nu}}\quad,
\end{displaymath}
and
\begin{displaymath}
\delta(\sqrt{-g}\,p)=\left(-\frac{1}{2}pg_{\sigma\nu}+\frac{\partial{p}}{\partial{\mu}}\frac{\partial{\mu}}{\partial{g^{\sigma\nu}}}\right)\sqrt{-g}\,\delta{g^{\sigma\nu}}\quad,
\end{displaymath}
but taking into account equation (\ref{142}) we have
\begin{displaymath}
\frac{\partial\mu}{\partial{g^{\sigma\nu}}}=-\frac{\mu}{2}U_{\sigma}U_{\nu}\quad.
\end{displaymath}
Furthermore, from equation (\ref{press}), we obtain
\begin{displaymath}
\frac{\partial{p}}{\partial{\mu}}=\frac{(\rho+p)}{\mu}\quad,
\end{displaymath}
so
\begin{displaymath}
\frac{\delta(\sqrt{-g}p)}{\delta{g^{\sigma\nu}}}=-\frac{\sqrt{-g}}{2}\left[pg_{\sigma\nu}+(\rho+p)\mu\,U_\sigma\,U_\nu\right]\quad.
\end{displaymath}
If $\frac{\delta{\mathcal{S}}}{\delta{g}}=0$, we find the well-known Einstein's equation for a perfect fluid
\begin{equation}
\frac{\delta{\mathcal{S}}}{\delta{g^{\mu\nu}}}=0\;\rightarrow\;R_{\sigma\nu}-\frac{1}{2}g_{\sigma\nu}R=8\pi[pg_{\sigma\nu}+(\rho+p)\,U_\sigma\,U_\nu]\quad.\label{150}
\end{equation}
By varying the action with respect to the velocity potentials $\beta$, $\alpha$, $S$ and $\theta$, respectively,  in an analogous procedure to that done previously, we found four non-linear coupled first order equations
\begin{eqnarray}
U^\nu\alpha_{,\nu}&=&0\quad,\label{152} \\
U^\nu\beta_{,\nu}&=&0\quad,\label{153} \\
U^\nu\theta_{,\nu}&=&T\quad,\label{154} \\
U^\nu{s}_{,\nu}&=&0\quad.\label{155}
\end{eqnarray}
Contracting (\ref{141}) with $U^{\nu}$ and using (\ref{153}) and (\ref{155}) we obtain
\begin{equation}\label{151}
U^\nu\phi_{,\nu}=-\mu\quad.
\end{equation}
Finally, when we vary the action with respect to $\phi$, we regain the baryon number conservation (\ref{130}).
\par
The equations (\ref{154}) and (\ref{151}) reflect the dynamic character of the potential $\theta$ and $\phi$ with their evolutions determined by the thermodynamic conditions of the fluid.
\par
The equations in the new formalism are also equivalent to those in the old formalism, as in the classical case. This proof of this equivalence is done in \cite{Schutz1} in a very elegant way.

\section{The Schutz's formalism in quantum cosmology}
\label{QCSF}


Quantum cosmology \cite{moniz1,Wiltshire,halliwell} is one of the least explored aspects of quantum gravity program. It is the application of quantum theory to the universe as a whole. This can be done because in the so-called Planck time, about $10^{-44}$s after the beginning, the universe had a size of $10^{-33}$cm which means that in this scale the quantum effects were dominant.
\par
In general the quantum cosmological models are derived by considering a finite number of fields and the presence of geometrical symmetry inherent to the universe, many of which are proven observationally. The quantization of a system subjected to these conditions is called quantization in a minisuperspace.
\par
Considering the early universe as the best laboratories where different quantum theories of gravitation can be tested, quantum cosmology has as one of their motivations to establish application limits to other theories and thus serve as a powerful auxiliary tool in getting a final theory of quantum gravity.
\par
From a purely cosmological point of view quantum cosmology plays an important role in solving a serious problem of the standard cosmological model, that is the existence of an initial singularity. Mechanisms such as quantum tunneling whose use is common in more conventional cosmological models \cite{halliwell} or even more sophisticated approaches like the loop quantum gravity where the volume is quantized and the evolution of the universe takes place in discrete time intervals \cite{isham, Rovelli, bojowald} are examples of consistent solutions of this problem.
\par
In addition, the theory allows us to establish the conditions for inflation, for the primordial perturbations and the seeds of the large scale structures of the universe and spontaneous symmetry breaking that must have happened in the early universe \cite{lawrie}. That is, the quantum cosmology can be considered as a theory of initial conditions.
\par
In this scenario, the Schutz's formalism is especially adequate. It has the advantage of ascribing dynamical degrees of freedom to the fluid to introduce the variable time. 

As an example, we analyze a perfect fluid Friedmann-Lemaitre-Robertson-Walker (FLRW) quantum cosmological model. The metric that describes such homogeneous and isotropic universe is 
\begin{equation}
\label{metri}
ds^2=-N(t)^2 dt^2 + a(t)\left(\frac{1}{1-kr^2}+r^2d\theta^2+r^2sin^2\theta d\phi^2\right)\quad,
\end{equation}
where $N(t)$ is the lapse function, which corresponds to the normal component of the four-velocity, $a(t)$ is the scale factor and $k$ is the spatial curvature in any time-slice of the universe, represented by three choices for the value of the parameter $-1$, $0$ e $1$. From the action of gravitation, without the presence of additional fields, given by
\begin{equation}
{\mathcal S}_g=\int d^4x\sqrt{-g}R\quad,\label{acao}
\end{equation}
we can use the Hamiltonian formalism to describe the system. With the equations (\ref{metri}) and (\ref{acao}), we obtain the Hamiltonian associated with the gravitational field, given by
\begin{equation}
H_g=-\frac{p_a^2}{24a}-6ka\quad,\label{hg}
\end{equation}
where $p_a$ is the canonical momentum associated to scale factor $a$. The above procedure is known as Hamiltonian formulation of general relativity or ADM formalism \cite{ADM}.  
\par
The matter content of the universe is modeled as a perfect fluid described by Schutz's formalism. In the case of a universe with FLRW metric the action of the fluid, given by
\begin{displaymath}
{\mathcal S}_f=\int\,d^4x\sqrt{-g}p\quad,
\end{displaymath}
can be rewritten as
\begin{displaymath}
{\mathcal S}_f=\int\,dt\,Na^3\rho\quad,
\end{displaymath}
where an overall factor of the spatial integral has been discarded, since it has no effect on the equations of motion. 
\par
Using the metric (\ref{metri}) and the energy density in terms of the enthalpy (\ref{densidade1}) (see details in appendix A), we have
\begin{equation}
\label{Sf}
{\mathcal S}_f=\int\,dt\,\left[Na^3(\gamma -1)\left(\frac{\mu}{\gamma}\right)^{\frac{\gamma}{\gamma - 1}}\,e^{-\frac{s}{\gamma -1}}\right]\quad.
\end{equation}
\par
As a consequence of the equation (\ref{142}) we have
\begin{equation}
\label{muu}
 \mu=\frac{1}{N}(\dot{\phi}+\theta\dot{s})\quad.
\end{equation}
Here we consider that the scalar potentials which describe rotational motion, are zero in the FLRW model because of its symmetry. 
Inserting above equation into equation (\ref{Sf}) we obtain
\begin{displaymath}
{\mathcal{S}}_f=\int\,dt\,\left[N^{-\frac{1}{\gamma - 1}}\,a^3\,\frac{\gamma - 1}{\gamma^{\frac{\gamma}{\gamma - 1}}}\,(\dot{\phi}+
\theta\dot{s})^{\frac{\gamma}{\gamma - 1}}\,e^{-\frac{s}{\gamma - 1}}\right]\quad,
\end{displaymath}
and therefore
\begin{equation}
\mathcal{L}_f=N^{-\frac{1}{\gamma - 1}}\,a^3\,\frac{\gamma - 1}{\gamma^{\frac{\gamma}{\gamma - 1}}}\,(\dot{\phi}+
\theta\dot{s})^{\frac{\gamma}{\gamma - 1}}\,e^{-\frac{s}{\gamma - 1}}\quad.\label{166}
\end{equation}
Thus, the canonically conjugate moments $\phi$ and $s$ can be rewritten as
\begin{equation}
P_{\phi}=N^{-\frac{1}{\gamma - 1}}\,a^3\,(\dot{\phi}+
\theta\dot{s})^{\frac{1}{\gamma -1}}\,\gamma^{-\frac{1}{\gamma -1}}\,e^{-\frac{s}{\gamma -1}}\quad,\label{167}
\end{equation}
and
\begin{displaymath}
P_s=\theta\,P_{\phi}\;.
\end{displaymath}

By canonical methods the Hamiltonian of the fluid $H_f$ can be expressed as
\begin{displaymath}
NH_{f} =  \dot{\phi}p_{\phi} + \dot{s}p_{s} - \mathcal{L}_f
\end{displaymath}
\begin{equation}
=(\dot{\phi}+\theta\dot{s})P_\phi-N^{-\frac{1}{\alpha}}a^3\frac{\alpha}{(\alpha+1)^{1+
\frac{1}{\alpha}}}(\dot{\phi}+\theta\dot{s})^{1+\frac{1}{\alpha}}e^{-\frac{s}{\alpha}}\quad.\label{169}
\end{equation}
Using equation (\ref{167}) we obtain
\begin{displaymath}
H_f=\frac{P_{\phi}^{\gamma}e^S}{a^{3(\gamma -1)}}\quad.\label{169.1}
\end{displaymath}
\par
To rewrite the Hamiltonian in a more simple way, we introduce the following canonical transformation
\begin{equation}
t = -P_s e^{-s}P_\epsilon^{-\gamma} \quad , \quad p_t =
P_\epsilon^{\gamma}e^s \quad , \quad \bar\epsilon = \epsilon -
\gamma\frac{P_s}{P_\epsilon} \quad , \quad \bar P_\epsilon =
P_\epsilon\quad,\label{170}
\end{equation}
so
\begin{equation}
H_f=\frac{p_t}{a^{3(\gamma -1)}}\quad,\label{171}
\end{equation}
where a linear canonical moment appears on the theory associated with the degrees of freedom of the perfect fluid.
\par
With this matter Hamiltonian, we can write the total Hamiltonian, $H$, which describes the dynamic behavior of the universe
\begin{equation}
H=H_g+H_f =
-\frac{p_a^2}{24a}-6ka+\frac{p_t}{a^{3(\gamma - 1)}}\quad.\label{178}
\end{equation}
According to Dirac’s quantization procedure we have
\begin{displaymath}
p_k\rightarrow\,\hat{p}_k=\pm\,i\frac{\partial}{\partial{q_k}}\quad,
\end{displaymath}
where $q_k$ are the generalized coordinates of the system and $p_k$ are his canonically associated moments. So
\begin{equation}
\hat{H}=\frac{\partial^2}{\partial{a^2}}-144ka^2+24\imath a^{(4-3\gamma)}\frac{\partial}{\partial{t}}\quad.\label{operador}
\end{equation}
\par
The fundamental equation of quantum cosmology is known as the Wheeler-DeWitt equation \cite{Wiltshire, halliwell}. This equation is obtained from the quantization of the canonical moments involved in the construction of the Hamiltonian formalism of general relativity and other moments of the theory. In the current context, this equation can be written making the equation (\ref{operador}) as an operator that annihilates the wave function of the universe $\Psi$, so that
\begin{equation}
\hat{H}\Psi(a,t)=\left[\frac{1}{24a}\frac{\partial^2}{\partial{a^2}}-6ka+\imath a^{(4-3\gamma)}\frac{\partial}{\partial{t}}\right]\Psi(a,t)=0\quad.\label{179}
\end{equation}
\par
We note that the introduction of degrees of freedom associated with the fluid leads to the appearance of a linear conjugated moment in this equation transforming it into a real Schr\"odinger equation with which we can describe the time evolution of the system. At this point the Schutz's formalism appears as a powerful tool to assign the degrees of freedom of the fluid, including its entropy, the crucial role of time.  \par Different cosmological scenarios can be studied by varying the fluid considered by choosing different values of $\gamma$ \cite{flavio1, pinto40, nivaldo1}, models with two perfect fluids, exotic matter and modified gravity $f(R)$ \cite{tovar2, flu, fracalossi, pedram, boseeinstein, vakili}. 
 The Schutz's formalism is used even with success in anisotropic quantum cosmology \cite{flavioaniso, flavioaniso2, pal1, pal2, pal3, pal4, pal5}, bouncing cosmologies \cite{pinto10, pinto20, pinto30} or very recently on Brans-Dicke theory and scalar-tensor theories \cite{flavioescalar, fabris2, fabris3, pal6}. 
\par Recently, the Seliger-Whitham and Schutz's formalism concepts have been applied in other proposals of reformulation of the action principle of the non-relativistic and relativistic perfect fluid \cite{ariki}.

\section{Final Words}
\label{Con}
In this paper, we have analyzed the construction of the hydrodynamical theory by means of variational principle making use of  velocity potentials in the classical and general relativistic cases. 
\par
For a perfect fluid in general relativity, we work with six velocity potentials. Variation of the Schutz's action with respect to the metric tensor yields Einstein's equations. Variation with respect to the velocity potencials reproduces the Eulerian equations of motion. Using this relativistic formalism, we obtained a consistent alternative to the problem of the absence of a time variable in the Wheeler-DeWitt equation in the quantum cosmology. 
\par
Our main care was being pedagogically clair. This is an important subject that is little known. As far as we know there is no a complete revision in the literature despite being widely used in quantum cosmology. We hope that this work can be useful for students and researches in relativistic hydrodynamics and quantum cosmology.

\newpage

\vspace{0.5cm} \noindent \textbf{Acknowledgments} \newline
\newline
\noindent
This work has received financial supporting from  FAPES (Brazil). Special thanks to Sridip Pal by suggested references.

\begin{appendices}

\section{Elementary aspects of the thermodynamics of one  perfect fluid}
\label{app}

Consider an one component perfect fluid composed by baryons, in the sense discussed by Fermi in his book \cite{fermi}. Since baryons may suffer transmutation, the pure mass of a group of baryons is not conserved. This situation does not happen with the number of baryons $N$. Thus, $m_HN$ is defined as the (conserved) rest mass  in a sample of matter containing $N$ baryons, where $m_H$ is the mass of the Hydrogen atom at the ground state. The difference between the total mass $E$ and $m_HN$ is called internal energy \footnote {in units of $c = 1$.} and is expressed by $U$. We denote by $\rho_0$ the rest mass density and by $u=\frac{U}{m_HN}$ the specific internal energy, both measured in a reference system momentarily at rest in fluid. This makes it possible to write
\begin{equation}
U=E-m_HN\quad,\label{111}
\end{equation}
so that
\begin{equation}
\label{energia1}
\frac{E}{m_HN}=u+1\quad.
\end{equation}
Now, by dividing the previous result by the volume of sample that contains $N$ baryons, you can write equation (\ref{energia1}) as
\begin{equation}
\rho=\rho_0(1+u)\quad.\label{114}
\end{equation}
\noindent where $\rho_{0}$ is the particle number density and $\rho$ is the energy density.
\par
We will use here a equation of state written as $p=p(\rho,u)$. We also will use the first law of thermodynamics
\begin{displaymath}
dW=dU+pdV\quad.
\end{displaymath}
By dividing the above equation by $m_HN$ we can write
\begin{equation}
dw=du+pd\left(\frac{1}{\rho_0}\right)\quad,\label{117}
\end{equation}
where $dw$ is the amount of energy per rest mass unit, added to the fluid in a quasi-static process.
\par
As the equation of state depends on two parameters, the Pfaff theorem \footnote{The Pfaff's theorem is an old theorem of differential forms. A good discussion can be found in Reference \cite{Schutz1}.} ensures the existence of functions such as specific entropy $s(\rho_0,u)$ and temperature $T(\rho_0,u)$, such that
\begin{equation}
du+pd\left(\frac{1}{\rho_0}\right)=Tds\quad.\label{118}
\end{equation}
\par
Then, we define inertial mass (enthalpy) \footnote{as defined and discussed in Reference \cite{Schutz1}} as
\begin{equation}
\mu=\frac{\rho+p}{\rho_0}=1+u+\frac{p}{\rho_0}\quad,\label{119}
\end{equation}
where the amount ($\rho+p$) behaves as inertial mass per unit volume of a perfect fluid. With this, we can use  $d\mu$ to eliminate $du$ in (\ref{118}), which allows to obtain
\begin{equation}
dp=\rho_0d\mu-\rho_0Tds\quad,\label{120}
\end{equation}
where we can see the possibility to express $\rho_0$ and $u$ as functions of $\mu$ and $s$ such that the equation of state takes the form $p=p(\mu,s)$. For the barotropic equation of state $p=(\gamma -1)\rho$, where $\gamma$ is a constant, we can write after some algebraic manipulations
\begin{displaymath}
(1+u)d[\ln(1+u)-(\gamma - 1)\,\ln\rho_0]=du+p d\left(\frac{1}{\rho_0}\right)=Tds\quad.
\end{displaymath}
It follows that
\begin{displaymath}
T=1+u\quad,
\end{displaymath}
and
\begin{displaymath}
s=\ln(1+u)-(\gamma -1)\,\ln\rho_0\quad.
\end{displaymath}
The last equation can be rewritten as
\begin{equation}
\rho_0=(1+u)^{\frac{1}{\gamma-1}}\,e^{-\frac{s}{\gamma-1}}\quad.\label{124}
\end{equation}
Furthermore, from equation (\ref{119}), we obtain
\begin{displaymath}
(1+u)=\frac{\mu}{\gamma}\quad.
\end{displaymath}
Replacing the previous equation in equation (\ref{124}), we can write
\begin{displaymath}
\rho_0=\left(\frac{\mu}{\gamma}\right)^{\frac{1}{\gamma-1}}e^{-\frac{s}{\gamma-1}}\quad,
\end{displaymath}
but
\begin{equation}
\label{densidade1}
\rho=\rho_0(1+u)=\rho_0\,\frac{\mu}{\gamma}=\left(\frac{\mu}{\gamma}\right)^{\frac{\gamma}{\gamma-1}}e^{-\frac{s}{\gamma-1}}\quad,
\end{equation}
and finally
\begin{equation}
\label{pressao1}
p=(\gamma -1)\left(\frac{\mu}{\gamma}\right)^{\frac{\gamma}{\gamma - 1}}e^{-\frac{s}{\gamma -1}}\quad.
\end{equation}
\noindent In the above expression the fluid pressure can be expressed directly in terms of their enthalpy.
\end{appendices}


\begin{thebibliography}{100}

\bibitem{landau} L. D. Landau and E. M. Lifshitz, Fluid Mechanics, Second Edition: Volume 6 (Course of Theoretical Physics S) 2nd Edition, Butterworth-Heinemann, (January 15, 1987).
\bibitem{Clebsch}
A. Clebsch,  J. Reine Angew.,  Math. 56, 1 (1859).
%
\bibitem{Seliger}
R. L. Seliger e G. B. Whitham, Proc. Roy. Soc. (London) A {\bf 305}, 1 (1968).
\bibitem{Schutz1}
B.F. Schutz, Phys. Rev.D{\bf 2}, 2762(1970).
%
\bibitem{Schutz2}
B.F. Schutz, Phys. Rev. D{\bf 4}, 3559(1971).
%
\bibitem{herivel} J. W. Herivel,  Proc. Camb. Phil. Soc. {\bf 51}, 344 (1955).
%
\bibitem{Lin} C. C. Lin, {\it Liquid  Helium}, in Proc.  Int.  School  Phys.  XXI (Academic Press (1963)).
%
\bibitem{Saarloos} W. Van Sarloos, D. Bedeaux and P. Mazur, Physica {\bf 107A}, 109 (1981).
%
\bibitem{HR} L. Smarr and C. Taubes, {\it General Relativistic Hydrodynamics: The Comoving, Eulerian, and Velocity Potential Formalisms}, in {\it Essays in General Relativity - A Festschrift for Abraham Taub}, edited by Frank J. Tipler, Academic Press (1980).
%
\bibitem{Tovar} F. T. Falciano, {\it Quantização do Modelo de Minisuperespa\c{c}o de Friedmann-Robertson-Walker Permeado por Poeira e Radiação via Interpreta\c{c}\~ao Causal da Mec\^anica Qu\^antica}, Disserta\c{c}\~ao de Mestrado, Centro Brasileiro de Pesquisas F\'{\i}sicas (2004).
%
\bibitem{moniz1} P. V. Moniz  {\it Quantum Cosmology - The Supersymmetric Perspective - vol. 1: Fundamentals (Lecture Notes in Physics)} (Springer 2010).

\bibitem{Wiltshire} D. L. Wiltshire, in B. Robson, N. Visvanathan and W.S. Woolcock (eds.), {\it Cosmology: The Physics of the Universe}, Proceedings of the 8th Physics Summer School, Australian National University, Canberra, Australia, 16 January – 3 February, 1995, (World Scientific, Singapore, 1996),  pp. 473-531, arXiv:gr-qc/0101003v2.

%
\bibitem{halliwell} J. J. Halliwell, in {\it Quantum cosmology and baby universes}, edited by S. Coleman, J. B. Hartle, T. Piran and S. Weinberg, Proceedings of the 7th Jerusalem Winter School for Theoretical Physics Jerusalem, Israel,  World Scientific Pub. Co. Inc. (1990).
%
\bibitem{isham} C.J. Isham, {\it Canonical quantum gravity and the problem of time}, arXiv:gr-qc/9210011v1.
%
\bibitem{Rovelli} C. Rovelli, {\it Quantum Gravity}, Cambridge Monographs on Mathematical Physics, (2004).
%
\bibitem{bojowald} M. Bojowald, Phys.Rev.Lett. {\bf 86},  5227 (2001).
%
\bibitem{lawrie} I. D. Lawrie,  Nuclear Physics B, Vol. {\bf 301}, Issue 4,  685 (1988).
%
\bibitem{ADM} R. Arnowitt, S. Deser and C.W. Misner, in L. Witten, ed., {\it{Gravitation: An Introduction to Current Reserch}}, Wiley, New York, 227-265 (1962).
%
\bibitem{flavio1} F.G. Alvarenga and N.A. Lemos, Gen. Rel. Grav. {\bf 30}, 681(1998),  arXiv:gr-qc/9802029v1.

\bibitem{pinto40} J. C. A. de Barros, N. Pinto-Neto and M. A. Sagioro-Leal, Phys. Lett. A {\bf 241}, 229 (1998)  arXiv:gr-qc/9710084v1.
%
\bibitem{nivaldo1} F.G. Alvarenga, J.C. Fabris, N.A. Lemos and G.A. Monerat, Gen. Rel. Grav. {\bf 34}, 651(2002), arXiv:gr-qc/0106051v2.

\bibitem{tovar2} N. Pinto-Neto, E. S. Santini and F. T. Falciano, Physics Letters A, {\bf 344}, Issue 2-4, 131 (2005).

\bibitem{flu} F.G. Alvarenga, R. Fracalossi, R. C. Freitas and S. V. B. Gon\c calves, {\it Classical and quantum cosmology with two perfect fluids: stiff matter and radiation}, arXiv:gr-qc/1607.03478v1.

%
\bibitem{fracalossi} G. A. Monerat, G. Oliveira-Neto, E. V. Corr\^ea Silva, L. G. Ferreira Filho, P. Romildo Jr., J. C. Fabris, R. Fracalossi, F. G. Alvarenga and S. V. B. Gon\c calves, Physical Review D, {\bf 76}, 024017 (2007), arXiv:gr-qc/0704.2585v1.
%
\bibitem{pedram} P. Pedram and S. Jalalzadeh, Physics Letters B, {\bf 659}, 6 (2008).

\bibitem{boseeinstein} F. G. Alvarenga, L. G. Ferreira Filho, R. Fracalossi, R. C. Freitas, S. V. B. Gon\c calves, G. A. Monerat, G. Oliveira-Neto and E. V. Corr\^ea Silva, {\it Primordial Universe with radiation and Bose-Einstein condensate}, arXiv:gr-qc/1610.06416v1.

\bibitem{vakili} B. Vakili, Class. Quant. Grav., {\bf 27}, 025008 (2010), arXiv:gr-qc/0908.0998v1.

\bibitem{flavioaniso} F. G. Alvarenga, A. B. Batista, J. C. Fabris and S. V. B. Gon\c calves,  Gen. Rel. Grav. {\bf 35} 1659 (2003), 
arXiv:gr-qc/:0304078v1.

\bibitem{flavioaniso2} F.G. Alvarenga, R. Fracalossi, R. C. Freitas and S. V. B. Gon\c calves, {\it The Kantowski- Sachs quantum model with stiff matter fluid}, arXiv:gr-qc/1506.02495v2.

\bibitem{pal1} S. Pal and N. Banerjee,  Phys. Rev. D {\bf 90} no.10, 104001 (2014), arXiv:gr-qc/1410.2718v1.

\bibitem{pal2} S. Pal and N. Banerjee,  Phys. Rev. D {\bf 91} no.4, 044042 (2014), arXiv:gr-qc/1411.1167v1.

\bibitem{pal3} S. Pal and N. Banerjee, Class.Quant.Grav. {\bf 32} no.20, 205005 (2015), arXiv:gr-qc/1506.02770v1.

\bibitem{pal4} S. Pal, Class.Quant.Grav. {\bf 33} no.4, 045007 (2016), arXiv:gr-qc/1504.02912v1.

\bibitem{pal5} S. Pal and N. Banerjee, {\it Anisotropic Models are unitary: A rejuvenation of standard quantum cosmology}, arXiv:gr-qc/1601.00460v1.

\bibitem{pinto10} N. Pinto-Neto, G. B. Santos and W. Struyve, Phys. Rev. D {\bf 89}, 023517 (2014)  arXiv:gr-qc/1309.2670v1.

\bibitem{pinto20} F. T. Falciano, N. Pinto-Neto and W. Struyve, Phys. Rev. D {\bf 91}, 043524 (2015)  arXiv:gr-qc/1501.04181v1.

\bibitem{pinto30} N. Pinto-Neto and J. C. Fabris, {\it Quantum Cosmology from the de Broglie-Bohm Perspective}, arXiv:gr-qc/1306.0820v1.

\bibitem{flavioescalar} F. G. Alvarenga, A. B. Batista and J. C. Fabris,  Int. J. Mod. Phys. {\bf D14}, 291 (2005), arXiv:gr-qc/0404034v1.

\bibitem{fabris2} C. R. Almeida, A. B. Batista, J. C. Fabris and P. R. L. V. Moniz,  Grav. Cosm. {\bf 21} no.3, 191 (2015), arXiv:gr-qc/1501.04170v1.

\bibitem{fabris3} C. R. Almeida, A. B. Batista, J. C. Fabris and P. R. L. V. Moniz, {\it Quantum Cosmology of scalar-tensor theories and self-adjointness}, arXiv:gr-qc/1608.08971v1. 

\bibitem{pal6} S. Pal, Phys.Rev. D {\bf 94}, 084023 (2016), arXiv:gr-qc/1608.06946v3.

\bibitem{ariki} T. Ariki and P. A. Morales, {\it Field Theory of the Eulerian Perfect Fluid}, arXiv:hep-th/1603.05935v3.

\bibitem{fermi} E. Fermi, Thermodynamics (Dover, New York, 1936), p. 91.

\end{thebibliography}
\end{document}